\pdfoutput=1
\documentclass{article}

\usepackage[preprint]{neurips_2023}

\usepackage[utf8]{inputenc} 
\usepackage[T1]{fontenc}    
\usepackage{hyperref}       
\usepackage{url}            
\usepackage{booktabs}       
\usepackage{amsfonts}       
\usepackage{nicefrac}       
\usepackage{microtype}      
\usepackage[table,xcdraw]{xcolor}         

\usepackage{subcaption}
\usepackage{wrapfig}
\usepackage{multirow}
\usepackage{makecell}
\usepackage{xspace}
\usepackage{enumitem}
\usepackage{amsmath}
\usepackage{listings}
\usepackage{algorithm}
\usepackage{algorithmic}
\usepackage{amsfonts,amssymb}
\usepackage{mathrsfs}
\usepackage{subcaption}
\usepackage{natbib}
\setcitestyle{numbers,square}
\usepackage{graphicx}       
\usepackage{cleveref}
\usepackage{fancyvrb}

\newcommand{\hhide}[1]{}
\newcommand{\hide}[1]{}
\newcommand{\model}[0]{\textsc{AutoGLM}\xspace}
\newcommand{\vpara}[1]{\vspace{0.07in}\noindent\textbf{#1}\xspace} %

\title{AutoGLM: Autonomous Foundation Agents for GUIs}

 \author{
Xiao Liu$^{12}$, Bo Qin$^{1\dagger}$, Dongzhu Liang$^{1\dagger}$, Guang Dong$^{1\dagger}$, Hanyu Lai$^{12*\dagger}$, Hanchen Zhang$^{12*\dagger}$, \\
\textbf{Hanlin Zhao$^{1\dagger}$, Iat Long Iong$^{12*\dagger}$, Jiadai Sun$^{1\dagger}$, Jiaqi Wang$^{1\dagger}$, Junjie Gao$^{1\dagger}$, Junjun Shan$^{1\dagger}$,} \\
\textbf{Kangning Liu$^{1\dagger}$, Shudan Zhang$^{12*\dagger}$, Shuntian Yao$^{1\dagger}$, Siyi Cheng$^{1\dagger}$, Wentao Yao$^{12*\dagger}$, }\\
\textbf{Wenyi Zhao$^{1\dagger}$, Xinghan Liu$^{12*\dagger}$, Xinyi Liu$^{1\dagger}$, Xinying Chen$^{1\dagger}$, Xinyue Yang$^{1\dagger}$, Yang Yang$^{1\dagger}$,}\\
\textbf{Yifan Xu$^{12*\dagger}$, Yu Yang$^{1\dagger}$, Yujia Wang$^{1\dagger}$, Yulin Xu$^{1\dagger}$, Zehan Qi$^{12*\dagger}$, Yuxiao Dong$^2$, Jie Tang$^2$} \\\\
Project Page: \url{https://xiao9905.github.io/AutoGLM}\\\\
Zhipu AI$^1$ \qquad Tsinghua University$^2$\\
 \\
{\includegraphics[height=3.5ex]{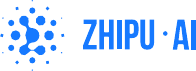}}
 }

\begin{document}

\maketitle
\renewcommand{\thefootnote}{\fnsymbol{footnote}}
    \footnotetext[1]{Work done while these authors interned at Zhipu AI.}
    \footnotetext[2]{These authors are listed alphabetically by first names.}
\renewcommand{\thefootnote}{\arabic{footnote}}

\vspace{-5mm}
\begin{abstract}
We present \model, a new series in the ChatGLM family~\cite{glm2024chatglm}, designed to serve as foundation agents for autonomous control of digital devices through Graphical User Interfaces (GUIs).
While foundation models excel at acquiring human knowledge, they often struggle with decision-making in dynamic real-world environments, limiting their progress toward artificial general intelligence.
This limitation underscores the importance of developing foundation agents capable of learning through autonomous environmental interactions by reinforcing existing models.
Focusing on Web Browser and Phone as representative GUI scenarios, we have developed \model as a practical foundation agent system for real-world GUI interactions.
Our approach integrates a comprehensive suite of techniques and infrastructures to create deployable agent systems suitable for user delivery.
Through this development, we have derived two key insights:
First, the design of an appropriate "intermediate interface" for GUI control is crucial, enabling the separation of planning and grounding behaviors, which require distinct optimization for flexibility and accuracy respectively.
Second, we have developed a novel progressive training framework that enables self-evolving online curriculum reinforcement learning for \model.
Our evaluations demonstrate \model's effectiveness across multiple domains.
For web browsing, \model achieves a 55.2\% success rate on VAB-WebArena-Lite (improving to 59.1\% with a second attempt) and 96.2\% on OpenTable evaluation tasks.
In Android device control, \model attains a 36.2\% success rate on AndroidLab (VAB-Mobile) and 89.7\% on common tasks in popular Chinese APPs.
Select \model capabilities are now available through the \href{https://chromewebstore.google.com/detail/\%E6\%99\%BA\%E8\%B0\%B1\%E6\%B8\%85\%E8\%A8\%80-\%E4\%BD\%A0\%E7\%9A\%84\%E6\%B5\%8F\%E8\%A7\%88\%E5\%99\%A8ai\%E5\%8A\%A9\%E6\%89\%8B/mnpdbmgpebfihcndnpgdaihnkmloclkd}{\color{blue}{Qingyan Browser Plugin}} for web applications and via \href{https://chatglm.cn/main/gdetail/6715f75ec8d0a702dff1e4e6}{\color{blue}{Form Applications}} for invited Android testing.
Additional results and materials will be released at \url{https://github.com/THUDM/AutoGLM}.
\end{abstract}

\section{Introduction}
Foundation models, including Large Language Models (LLMs)~\cite{GPT3,openai2023gpt,chowdhery2022palm,Claude-3,zeng2022glm,glm2024chatglm} and Large Multimodal Models (LMMs)~\cite{liu2024visual,GPT-4-vision-preview,GPT-4o,Claude-3.5-Sonnet}, have captured widespread attention for their remarkable language understanding and generation capabilities.
Through extensive self-supervised~\cite{liu2021self} pre-training on internet-scale corpora, these models have acquired not only knowledge and language abilities but also human-like reasoning and planning capabilities, giving rise to \textbf{LLMs as Agents}~\cite{liu2023agentbench,park2023generative}.
These agents have demonstrated their utility across diverse domains, including coding~\cite{wang2024opendevin,jimenez2023swe,zhang2024naturalcodebench}, data analysis~\cite{hu2024infiagent,liu2023agentbench}, and gaming~\cite{wang2023voyager,light2023avalonbench}, charting a promising course toward Artificial General Intelligence (AGI) through the development of multimodal \textbf{Foundation Agents}~\cite{liu2024visualagentbench} that serve as generalists across multiple tasks and environments.

\begin{figure}[t]
    \centering
    \includegraphics[width=\linewidth]{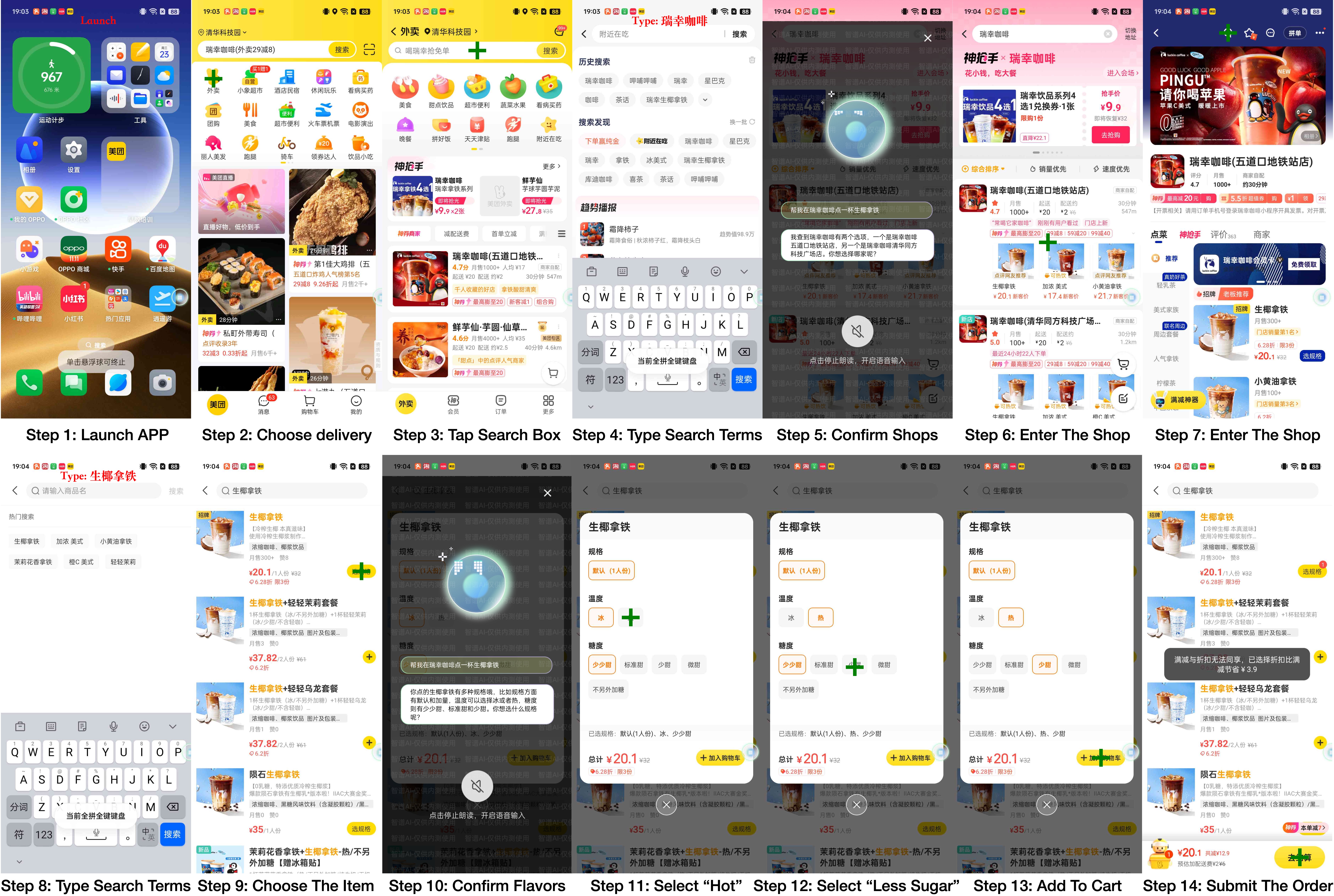}
    \caption{\model real phone use demonstration for instruction ``\texttt{Order a hot coconut latte from Luckin Coffee with half sugar}'' for Chinese Android APP Meituan. \href{https://xiao9905.github.io/AutoGLM}{\color{blue}{See more videos}}. }
    \label{fig:android_example}
    \vspace{3mm}
    \includegraphics[width=\linewidth]{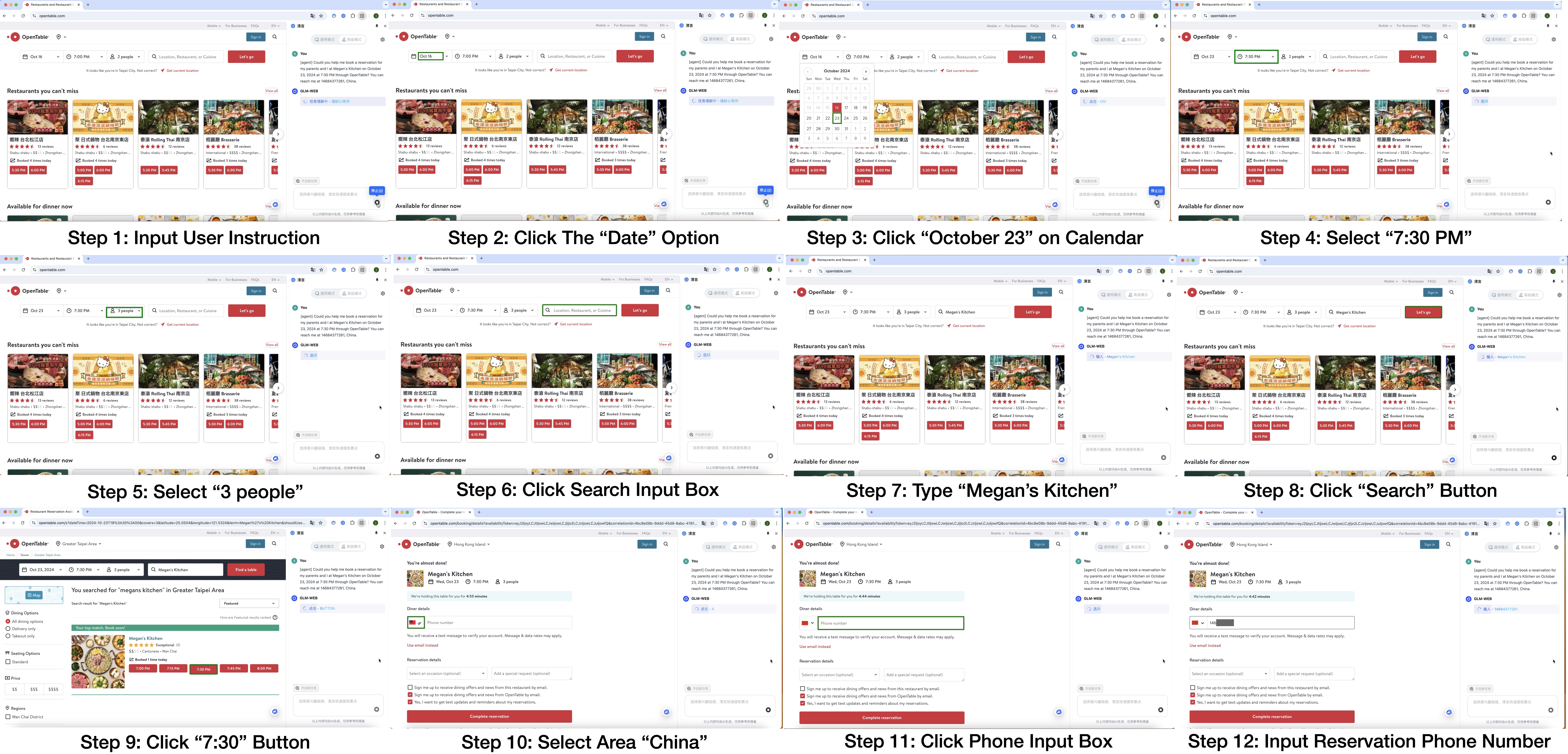}
    \caption{\model real web-browser use demonstration for instruction ``\texttt{Could you help me book a reservation for my parents and I at Megan's Kitchen on October 23, 2024 at 7:30 PM through OpenTable? You can reach me at 146xxxxxxxx, China.}'' on the website OpenTable. \href{https://xiao9905.github.io/AutoGLM}{\color{blue}{See more videos}}.}
    \label{fig:web_example}
\end{figure}

The ubiquity of digital devices presents a unique opportunity for GUI-capable agents~\cite{hong2023cogagent,zheng2024gpt,zhang2023appagent,lai2024autowebglm}. 
This domain offers several advantages: GUI simulators can be readily deployed in parallel for data annotation and online reinforcement learning (RL); GUI environments provide rich textual and visual inputs essential for foundation model agents, but in a safer and controllable environments compared to embodied environments; and GUI agents hold broad practical appeal given their extensive potential user base. Their successful development could fundamentally transform human-device interaction.

However, the development of foundation agents for GUI faces a critical challenge: the scarcity of decision-making data in existing pre-training sets. 
While the internet contains vast human knowledge, it primarily consists of \textit{static} information that inadequately captures human decision-making and environmental interaction. 
Building capable foundation agents requires enriching them with \textit{dynamic} knowledge, either through direct interaction with real-world environments or through learning from synthesized trajectories. 
Such foundation agents can then self-evolve in the digital world, iteratively improving to achieve genuine general intelligence.

Crucially, these systems must be developed with progressive user deployment in mind. 
Autonomous agents are designed to augment, not replace, human capabilities. 
User deployment serves the dual purpose of teaching agents effective human assistance while allowing humans to adapt to intelligent assistants. 
This approach also enables researchers to systematically understand, discover, and examine both the potential benefits and risks of autonomous foundation agents during development.

In response to these opportunities and challenges, we introduce \model, a series of foundation agents built upon the ChatGLM~\cite{glm2024chatglm} model family. 
\model represents a pioneering attempt to develop foundation agent prototypes for two fundamental GUI scenarios: Web Browser and Android. 
To address the data scarcity challenge, we employ a comprehensive suite of training techniques and develop key infrastructures for user deployment. 
This process has yielded two crucial insights:

\begin{itemize}[leftmargin=1.5em,itemsep=0pt,parsep=0.2em,topsep=0.0em,partopsep=0.0em]
\item \textbf{Intermediate Interface Design}: We find it essential to design an intermediate interface that disentangles planning and grounding behaviors in foundation GUI agents. They present distinct requirements – planning demands flexibility and error recovery, while grounding emphasizes action accuracy. Their separation enables more agile development and enhanced performance.
\item \textbf{Self-Evolving Online Curriculum RL~\cite{qi2024webrl}}: We recognize that error recovery~\cite{liu2024visualagentbench} is crucial for robust and deployable agent applications, yet it remains difficult to acquire through offline training alone. Additionally, the shortage of instructions and trajectories impedes training progress. We address this challenge through self-evolving RL, implemented according to a progressive weak-to-strong curriculum schedule in an online manner.
\end{itemize}

Building on these insights, \model demonstrates exceptional capabilities across various benchmarks and real-world tests. 
In Web Browsing, \model achieves a task success rate (SR) of 55.2\% on the challenging VAB-WebArena-Lite~\cite{zhou2023webarena,liu2024visualagentbench}, substantially surpassing GPT-4o's 18.2\%. 
With a second attempt opportunity, this improves to 59.1\%. On OpenTable real-world booking tasks, \model achieves 96.2\% SR, outperforming both GPT-4o (62.6\% SR) and Agent Q~\cite{putta2024agent} (81.7\%). 
Select \model web capabilities are publicly available via the \href{https://chromewebstore.google.com/detail/\%E6\%99\%BA\%E8\%B0\%B1\%E6\%B8\%85\%E8\%A8\%80-\%E4\%BD\%A0\%E7\%9A\%84\%E6\%B5\%8F\%E8\%A7\%88\%E5\%99\%A8ai\%E5\%8A\%A9\%E6\%89\%8B/mnpdbmgpebfihcndnpgdaihnkmloclkd}{\color{blue}{Qingyan Browser Plugin}} on both Chrome and Edge Plugin Store.
See the real example in Figure~\ref{fig:web_example}.

For Android control, \model achieves 36.2\% SR on AndroidLab~\cite{xu2024androidlab} (previously known as VAB-Mobile~\cite{liu2024visualagentbench}), a comprehensive interactive Android evaluation framework. 
This performance exceeds both GPT-4o (31.2\% SR) and Claude-3.5-Sonnet (29.0\% SR). We have also implemented a practical Android application via AccessibilityService for autonomous device control. 
In human evaluation, \model achieves an impressive 89.7\% SR on common tasks (e.g., ``\texttt{Please Order a large iced Americano with half sugar from the nearest coffee shop for delivery to my company}'') across popular Chinese APPs. 
The Android client is currently available for invited internal testing through \href{https://chatglm.cn/main/gdetail/6715f75ec8d0a702dff1e4e6}{\color{blue}{Form Applications}}.
See the real example in Figure~\ref{fig:android_example}.
\section{\model: Techniques and Insights}
In this section, we will give an overview of the techniques involved in developing \model.
Particularly, we will discuss the two important insights that enable \model's significant improvements compared to existing LLM or LMM-based GUI Agents.

\subsection{Important Techniques}
Training agents can be different from training ordinary LLMs or LMMs.
A key obstacle lies in the lack of high-quality trajectory data that entails the decision-making process.
Following are some useful techniques we realized during the project.

\vpara{Pre-training.}
Generally, there is little agent-related data on internet text corpora, making LLMs fall short of effectively act as agents.
Additionally, existing LMM pre-training, which is primarily ``visual instruction tuning'', models the alignment between texts and images without sufficiently learning from sequential multimodal data~\cite {baker2022video,fan2022minedojo}.
As a result, properly leveraging existing online data with weak-supervised decision-making signals in pre-training would actually help.
Besides, for multimodal perception, high-resolution visual inputs are very important according to CogAgent~\cite{hong2023cogagent} and our observations, especially when using grounding strategies like Set-of-Marks (SoM) prompting~\cite{yang2023set}.

\vpara{Large Multimodal Models (LMMs).}
LMMs are important for GUI understanding and manipulation.
Traditionally in Robotic Process Automation (RPA), the paradigm has been using Optical Character Recognition (OCR) catcher to match key elements in human handcrafted automating programs, which cannot be scaled and generalized.
LMMs, instead, can perform fuzzy matching and do long-horizon planning thanks to its strong grasping of commonsense and GUI environments from pre-training.
Nevertheless, LMMs still require much training to gain strong planning and reasoning abilities necessary for agent tasks.

\vpara{Behavior Cloning (Supervised Fine-tuning).}
Behavior Cloning (BC) is a key strategy for training agents from scratch with high-quality expert trajectories.
The strategy has also been verified as effective for LLM and LMM-based agent training~\cite{nakano2021webgpt,zeng2023agenttuning,chen2023fireact,hong2023cogagent,lai2024autowebglm,liu2024visualagentbench}.
Nevertheless, it is of extreme cost and time to collect expert trajectories.
Moreover, a fundamental problem of using BC is that agents only learn to imitate experts' behaviors step-by-step without fully understanding its goal.
When expert trajectories are oracle (mostly the case for maintaining training stability), agents fail to foster abilities to recover from errors well~\cite {liu2024visualagentbench}.

\vpara{Curriculum Learning.}
Agent tasks are usually of substantially varied difficulties.
As a result, it is wise to progressively add difficulty to training with a curriculum schedule.
For example, AutoWebGLM~\cite{lai2024autowebglm,iong2024openwebagent} adopts a multi-stage curriculum, where agent models are sequentially trained with single-step tasks, simple few-step tasks, and complicated long-horizon tasks.
DigiRL~\cite{bai2024digirl} also proposes a simple curriculum to filter appropriate tasks from a fixed set of instructions according to the corresponding agent capabilities at a certain timestamp.
We basically find the strategy very useful for building foundation agents with complex goal-achieving abilities.

\vpara{Reward Modeling (RM).}
To enable online RL with foundation agents, a proper RM is necessary for providing supervision.
Traditionally, many RL agents are trained with limited tasks with precise rule-based reward functions.
However, foundation agents based on LLMs and LMMs are targeting generalist mission accomplishment in open worlds, which contradicts task-specific reward functions' abilities.
Therefore, it is crucial to build generalizable RMs that can deal with a wide range of real-world agent tasks.
Specifically, RMs can be categorized to outcome-supervised ORM and process-supervised PRM~\cite{lightman2023let,creswell2022selection,zelikman2022star}, which provide different granularities of effective supervision.

\vpara{Reinforcement Learning (RL).}
Compared to BC, RL from a narrow sense can better learn from failures.
This is especially important for foundation agent training since high-quality expert trajectories are extraordinarily hard to acquire~\cite{nakano2021webgpt}.
However, the challenge in applying RL to foundation agent training lies in the inefficiency of sampling in environments.
The problem can be understood from two aspects:
1) Simulators: when agents are exploring in the Web or Android environments, their efficiency is bounded by internet connection speed and maximum degree of parallelism. Environments like Android Virtual Devices are quite memory-consuming~\cite{liu2024visualagentbench}.
2) Sample Diversity: LLMs and LMMs are trained to output certain function-based actions. The strict function formatting usually requires overfitting training with the model, resulting in stubborn monotonous sampling results even when they are inferenced with a high temperature~\cite{wang2024planning}. 

Despite the challenge, we believe to scale up RL and post-training on foundation models is crucial for building strong foundation agents, as indicated by the success of \texttt{OpenAI o1}.
It is unlikely to build general intelligence without letting it learn from interactions with real-world environments.

\subsection{Insight 1: Intermediate Interface Design}
During the development, we find intermediate interface design vital for disentangling the behaviors of planning and grounding in foundation agents.
By separating them into different modules, foundation agents can be improved from both dimensions of flexibility and accuracy without interference.

The intuition is simple: we find existing LLMs and LMMs to be more capable in planning than in grounding when executing agent tasks on existing benchmarks.
While the planning could still be significantly improved, a majority of current errors arise from incorrect element identification in the grounding period~\cite{liu2024visualagentbench}.
For example, a typical action generated in VAB-WebArena-Lite when testing with visual inputs would be:
\begin{Verbatim}[commandchars=\\\{\}]
\small{\textcolor{gray}{# End-to-End Agent}}
\small{\textcolor{blue}{do(action="Click", element_coordinates=[823,684])}}
\end{Verbatim}
where the element ``\texttt{4}'' may refer to a "Submit" button on Reddit.
If we change the format to the following:
\begin{Verbatim}[commandchars=\\\{\}]
\small{\textcolor{gray}{# Agent with Intermediate Interface Design}}
\small{\textcolor{gray}{# Planner}}
\small{\textcolor{blue}{do(action="Click", element_description="the `Submit' button on the bottom right")}}
\small{\textcolor{gray}{# Grounder}}
\small{\textcolor{blue}{find_coordinates_by_instruction("the `Submit' button on the bottom right")}}
\small{\textcolor{gray}{# Expected Output: [823,684]}}
\end{Verbatim}

\begin{table}[t]
\caption{Experiments on Intermediate Interface Design on VAB-WebArena-Lite~\cite{zhou2023webarena,liu2024visualagentbench}.}
\vspace{1mm}
\resizebox{\columnwidth}{!}{
\begin{tabular}{@{}lccc@{}}
\toprule
Observation Type              & \texttt{gpt-4o} (text)       & \texttt{gpt-4o} (visual)    & \texttt{gpt-4-vision-preview} (visual) \\ \midrule
End-to-End Agent              & 14.3\%                       & 18.2\%                      & 18.8\%                                 \\
Intermediate Interface Design & 18.1\% \color{red}{(+3.8\%)} & 27.3\% \color{red}{(+9.1\%)} & 36.4\% \color{red}{(+17.6\%)}           \\ \bottomrule
\end{tabular}
}
\label{tab:intermediate_interface_design}
\end{table}

In this way, planner and grounder abilities could be separately improved.
Actually, it is much easier to harvest massive grounding data from automatic construction from unsupervised environmental observations.
In our experiments (Cf. Table~\ref{tab:intermediate_interface_design}), we find the strategy with the grounder we trained to be very useful for improving proprietary LLM/LMM API-based planners.
Our observation is similar to another concurrent work~\cite{gou2024navigating} which explores a universal grounding model for GUI agents.

\subsection{Insight 2: Self-Evolving Online Curriculum RL}
While Intermediate Interface Design helps alleviate the issue of inaccurate grounding, planning is still a problem.
Many existing agent works in literature base their frameworks on proprietary LLM/LMM APIs, whose planning abilities consequently fail to be improved by training.

\begin{wrapfigure}{r}{0.48\textwidth}
    \centering
    \vspace{-5mm}
    \includegraphics[width=1\linewidth]{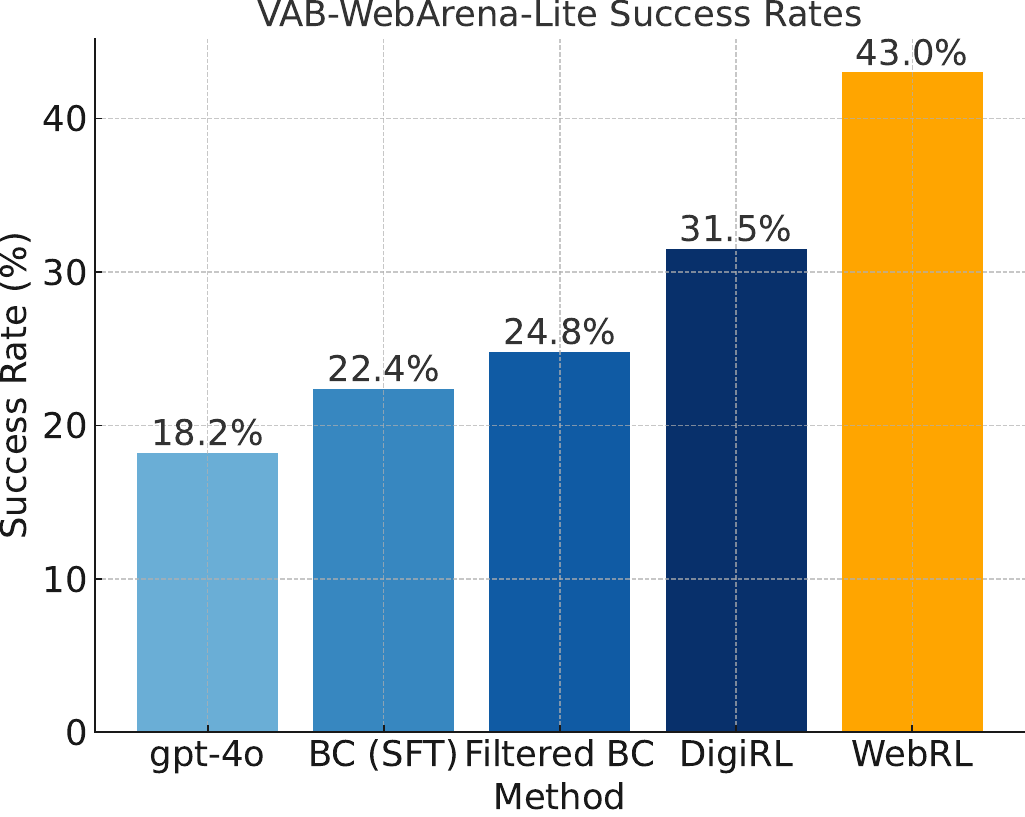}
    \vspace{-3mm}
    \caption{VAB-WebArena-Lite~\cite{zhou2023webarena,liu2024visualagentbench}, methods are all trained using GLM-4-9B-Base~\cite{glm2024chatglm}.}
    \label{fig:webrl}
    \vspace{-8mm}
\end{wrapfigure}

As a result, we decide to explore training in-house planners via RL.
It is very challenging as it lacks a sufficient amount of either user tasks or expert trajectories.
We develop a self-evolving online curriculum RL framework--WebRL~\cite{qi2024webrl}--for training foundation agents from scratch.
Take WebArena~\cite{zhou2023webarena} environment as an example, we adopt the actor-critic RL framework for training.
Briefly speaking, we identify the most difficult issues when we apply curriculum RL to the problem--task data scarcity and policy distribution drift.

\vpara{Task Data Scarcity.}
Leveraging around 1,000 BC data provided by VisualAgentBench~\cite{liu2024visualagentbench}, we initialize GLM-4-9B to 22.4\% SR.
At this point, we have run out of either task instructions or oracle trajectories.
Thus we apply self-evolving techniques to augment failed task instructions during online roll-outs, mutating instructions to be more complicated or simpler.
These self-evolved instructions are filtered by the critic and then used for roll-outs in the next iterative training phase.

\vpara{Policy Distribution Drift.}
One significant problem of curriculum learning is the policy distribution drift during the progressive curriculum schedule.
We develop a KL-constrained policy update for agent training, together with actor confidence filtered experience replay.
The ablation study indicates that the designs are indispensable for consistent performance improvement during iterative training.

\section{Results}
In this section, we report our evaluation of \model on both Web and Android-oriented tasks.

\begin{figure}[t]
  \centering
    \centering
    \includegraphics[width=0.9\linewidth]{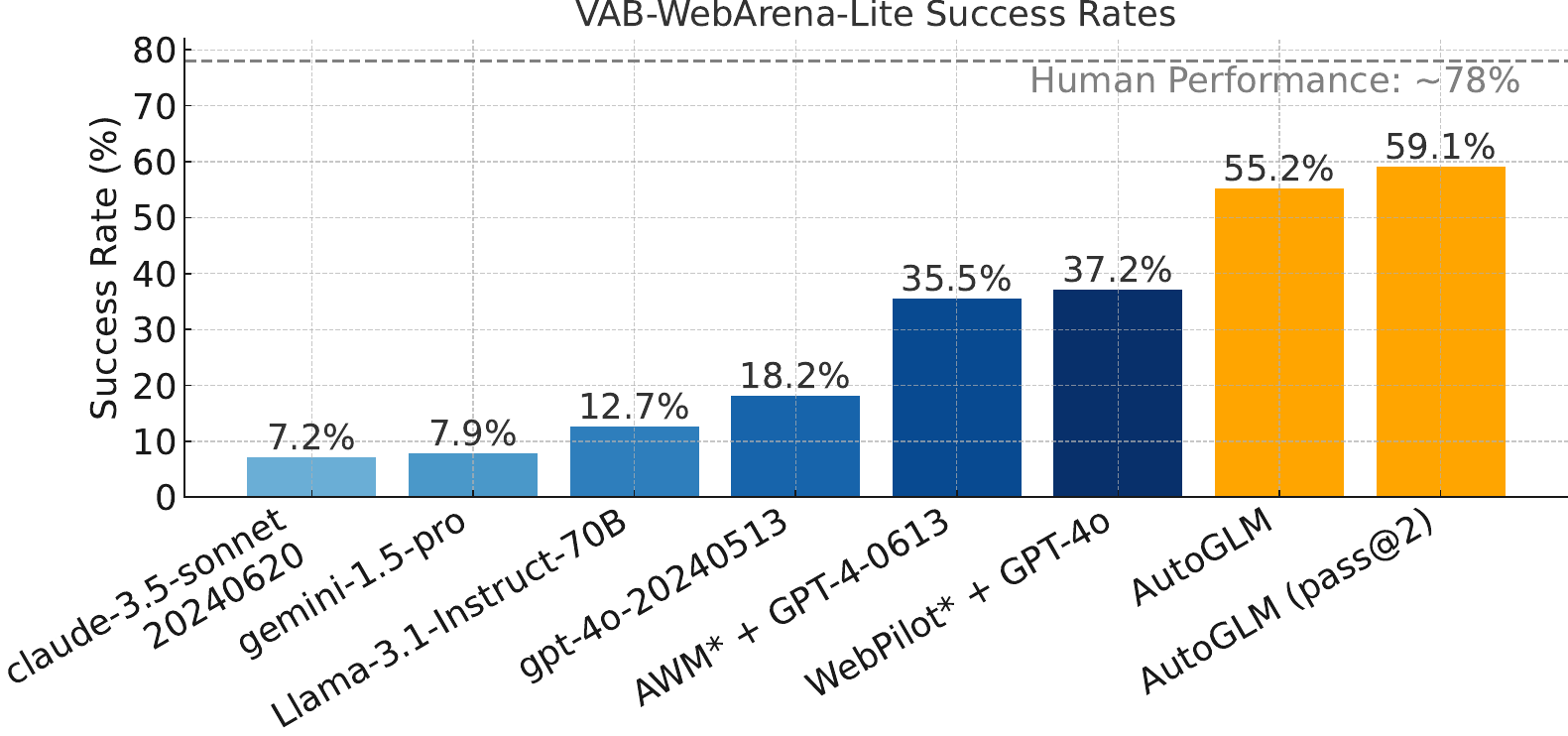}
    \caption{Success rates of different agents on VAB-WebArena-Lite. $^*$Results of AWM and WebPilot are evaluated on full WebArena and taken from reporting.}
    \label{fig:webarena}
  \vspace{-5mm}
\end{figure}

\subsection{Evaluated on Web}
We adopt three interactive benchmarks: VAB-WebArena-Lite~\cite{zhou2023webarena,liu2024visualagentbench} and an online human evaluation dataset OpenTable~\cite{putta2024agent}.
\model experiences training optimization in these environments.

\vpara{VAB-WebArena-Lite~\cite{zhou2023webarena,liu2024visualagentbench}.}
VAB-WebArena-Lite\footnote{\url{https://github.com/THUDM/VisualAgentBench/blob/main/VAB-WebArena-Lite}} is a refined 165-task subset of the original 812-task WebArena~\cite{zhou2023webarena} with manual verification of answers and judge functions.
Its design intention is to speed up the evaluation on WebArena and ensure judging correctness.
We evaluate representative proprietary LLM/LMM APIs, open models~\cite{dubey2024llama}, recent agent frameworks~\cite{wang2024agent,zhang2024webpilot}, and \model.
Results in Figure~\ref{fig:webarena} show that \model has significantly advanced on the benchmark, narrowing the performance gap between autonomous agents and humans.

\begin{wrapfigure}{r}{0.38\textwidth}
    \centering
    \vspace{-5mm}
    \includegraphics[width=1\linewidth]{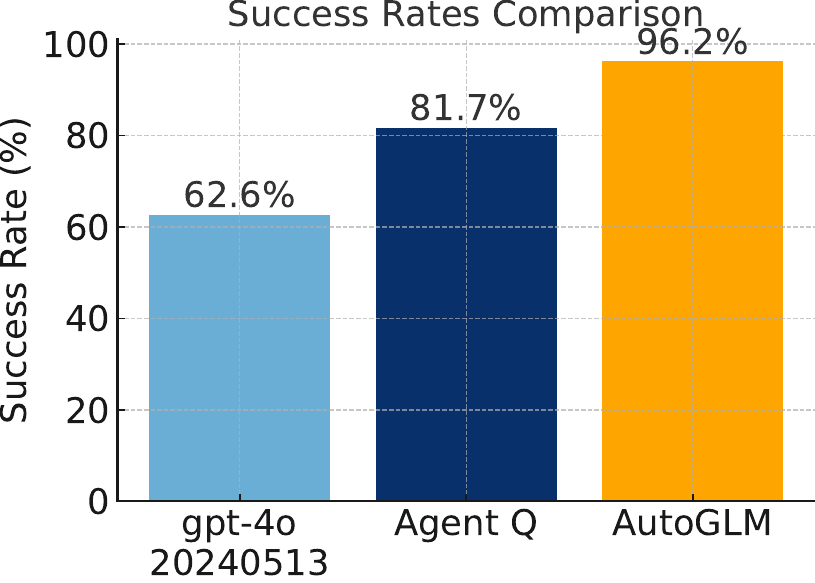}
    \vspace{-3mm}
    \caption{OpenTable Eval~\cite{putta2024agent}.}
    \label{fig:opentable}
    \vspace{-8mm}
\end{wrapfigure}

\vpara{OpenTable Eval~\cite{putta2024agent}.}
Following Agent Q~\cite{putta2024agent}, we also evaluate \model on a real website OpenTable, which provides an online open booking service.
Since the test set of~\cite{putta2024agent} is undisclosed, we reconstruct a 200-sample test set according to the example provided in its paper (``\texttt{Book reservation for restaurant Cecconi’s on OpenTable for 4 people on May 22 2024 at 7:00 PM}'') and run evaluation on the real OpenTable website with human evaluation.
Results are in Figure~\ref{fig:opentable}.
\model outperforms both gpt-4o and Agent Q on this real-world website.

\subsection{Evaluated on Android}
We evaluate \model's Android abilities on the academic benchmark AndroidLab~\cite{xu2024androidlab} (i.e., VAB-Mobile~\cite{liu2024visualagentbench}) and frequent tasks in common Chinese Mobile APPs on Android.

\begin{figure}[t]
  \centering
    \centering
    \includegraphics[width=0.9\linewidth]{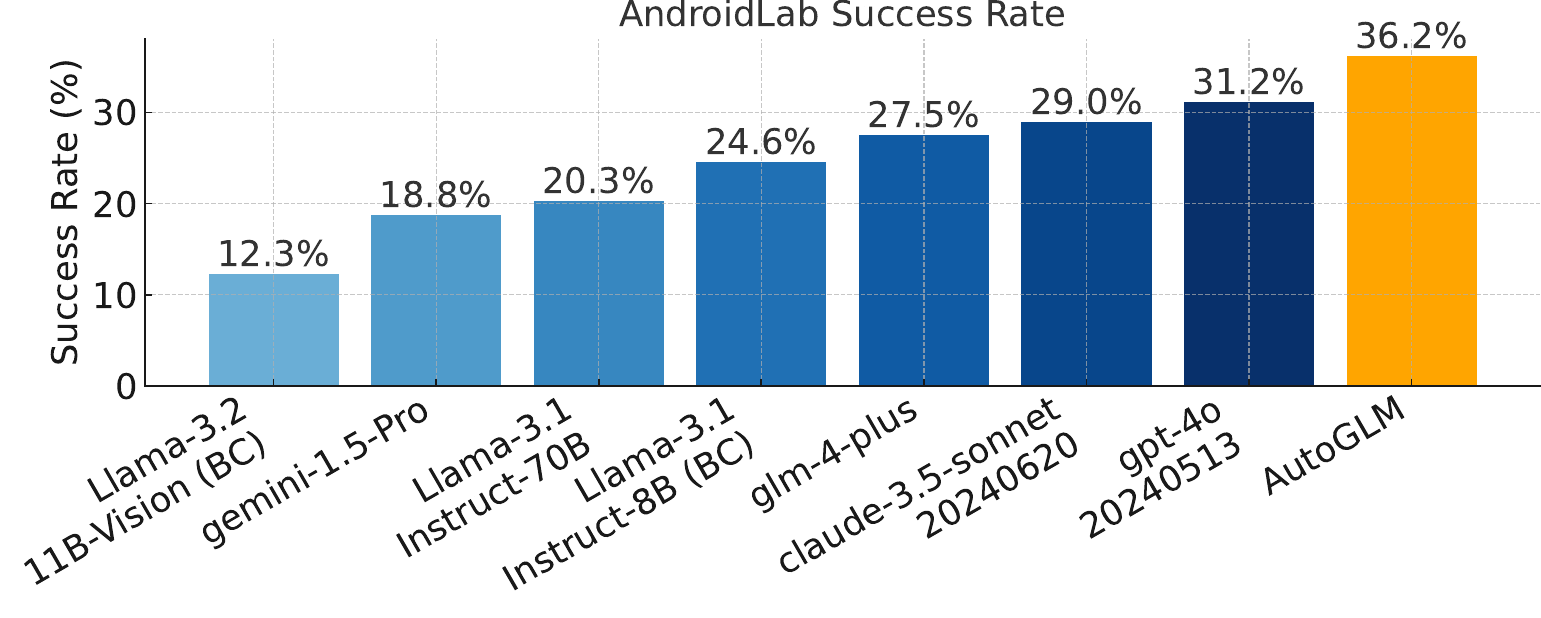}
    \caption{Success rates of different agents on AndroidLab. BC indicates behavior cloning training.}
    \label{fig:androidlab}
  \vspace{-5mm}
\end{figure}

\begin{wrapfigure}{r}{0.4\textwidth}
    \centering
    \vspace{-6mm}
    \includegraphics[width=1\linewidth]{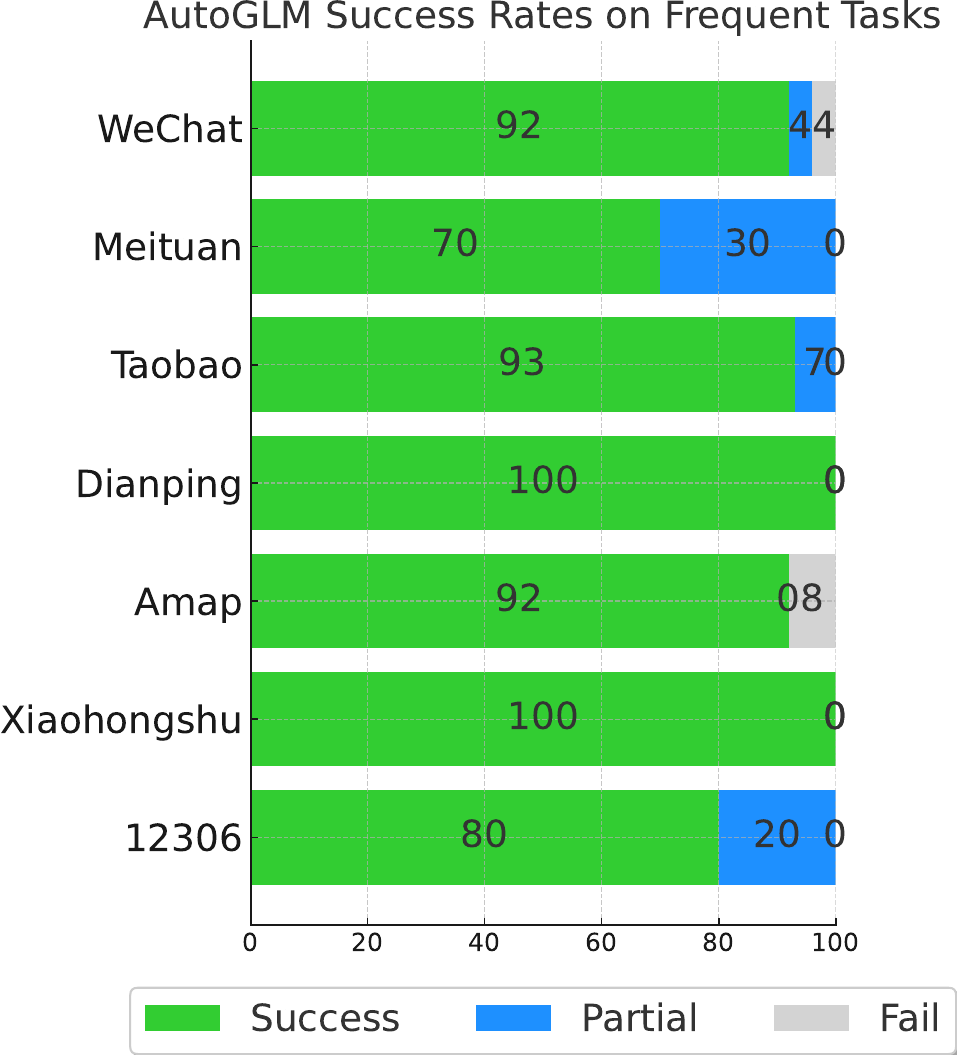}
    \vspace{-4mm}
    \caption{Human evaluated success rates of \model on Chinese APPs.}
    \label{fig:chinese_apps}
    \vspace{-10mm}
\end{wrapfigure}

\vpara{AndroidLab~\cite{xu2024androidlab} (VAB-Mobile~\cite{liu2024visualagentbench}).}
AndroidLab is an interactive Android benchmark and development environments that support reproducible evaluation, covering systems and some offline deployable English APPs.
Compared to some existing benchmarks such as AITW~\cite{rawles2024androidinthewild}, its interactive nature allows more practical evaluation of foundation agents for Android and improvement via RL.
We evaluate representative proprietary LLM/LMM APIs, open models~\cite{dubey2024llama} fine-tuned on provided BC data, and \model.
Results are shown in Figure~\ref{fig:androidlab}.
\model achieves 36.2\% SR, the best-performed one among all compared agents.

\vpara{Human Evaluation on Chinese Android APPs.}
To test the practicality of \model being deployed for public users, we carefully examine it on frequent tasks in 7 common Chinese Android APPs, including WeChat, Meituan, Taobao, Dianping, Amap, Xiaohongshu, and 12306.

\begin{table}[t]
\caption{Examples of test queries for evaluating \model on Chinese APPs.}
\vspace{1mm}
\renewcommand\arraystretch{1.2}
\resizebox{\columnwidth}{!}{
\begin{tabular}{@{}ll@{}}
\toprule
APP         & Test Query Example                                                                                                                  \\ \midrule
WeChat      & \texttt{Post a complimentary comment on Alice's latest post on Moments}                                                            \\
Meituan     & \texttt{Order a cold coconut latte from nearest coffee shop, with half sugar}                                                       \\
Taobao      & \texttt{Check the shipping/delivery status of the shirt I ordered}                                                                  \\
Dianping    & \texttt{What's the nearest restaurant from Beijing's must-eat list to my location}                                                  \\
Amap        & \texttt{Order a ride/taxi to Taikoo Li Sanlitun, immediately}                                                                                    \\
Xiaohongshu & \texttt{Bookmark and summarize the most liked travel guide post about Aranya} \\
12306       & \texttt{Book a bus ticket from Zhuhai to Guangzhou for Saturday}                                                                    \\ \bottomrule
\end{tabular}
}
\label{tab:example_queries}
\end{table}

We curate a test query set (Cf. Table~\ref{tab:example_queries}) for evaluating \model's real performance in the user delivery setting, where the final success rate is determined by human evaluation on the whole executing trajectories.
Instead of evaluating an Android Virtual Device (AVD) as in AndroidLab and previous work~\cite{rawles2024androidworld,yang2023appagent}, our evaluation is conducted in physical Android phones implemented with AccessibilityService applications to reflect the practical scenarios of foundation agents for phone use.
Results are shown in Figure~\ref{fig:chinese_apps}.
We classify results into 3 types for better understanding of \model:

\begin{itemize}[leftmargin=1.5em,itemsep=0pt,parsep=0.2em,topsep=0.0em,partopsep=0.0em]
\item \textbf{Success}: The task is completely successful, fulfilling all requirements in user instructions.
\item \textbf{Partial}: The task is partially done in the correct direction without completing some following procedures to fulfill user requirements.
\item \textbf{Fail}: The task is terminated too early, gets stuck in the middle, or goes in the wrong direction.
\end{itemize}

As we observe, \model works decently on evaluated APPs.
While currently, it is unable to solve all tasks perfectly, unfinished tasks are able to be half completed, which would still be of assistance in practical scenes to users to speed up GUI operations.

\section{Conclusion}
Through this work, we introduced \model, a series of foundation agents built upon the ChatGLM model family that demonstrates strong capabilities in GUI operation across web browsing and Android environments. Our key contributions include the design of an intermediate interface that effectively disentangles planning and grounding behaviors, and the development of a self-evolving online curriculum RL approach that enables robust error recovery and performance improvement. The strong empirical results across various benchmarks, including a 55.2\% success rate on VAB-WebArena-Lite and 36.2\% on AndroidLab, along with successful real-world deployments through browser plugins and Android applications, demonstrate \model's potential as a significant step toward developing practical foundation agents for GUI interaction.


\bibliographystyle{abbrv}
\bibliography{ref}

\end{document}